\documentclass[aps,prd,reprint,groupedaddress,amsmath,amssymb]{revtex4-1}
\usepackage[T1]{fontenc}
\usepackage{lmodern}
\usepackage{graphicx,mathrsfs,bm,mathtools,booktabs,array,enumitem}
\usepackage{color}
\usepackage[
colorlinks=true,
filecolor=black,
anchorcolor=blue,
linkcolor=blue,
citecolor=cyan,
urlcolor=cyan,
linktocpage=true,
plainpages=false,
breaklinks=true,
pdfstartview=FitH
]{hyperref}
\hypersetup{pdftitle={Weyl double copy for the relativistic Rindler fluids},pdfauthor={}}
\DeclareMathAlphabet{\mathpzc}{OT1}{pzc}{m}{it}
\newcommand{\dd}{\mathrm{d}}
\newcommand{\ii}{\mathrm{i}}
\newcommand{\Order}{\mathcal{O}}
\newcommand{\Kc}{\mathcal{K}}
\newcommand{\calB}{\mathcal{B}}
\newcommand{\KNR}{\mathcal{K}_{\mathrm{NR}}}
\newcommand{\NR}{\mathrm{NR}}
\newcommand{\rel}{\mathrm{rel}}
\allowdisplaybreaks[3]
\usepackage{etoolbox}
\makeatletter
\patchcmd{\l@@sections}
  {\rightskip\tocleft@pagenum plus 1fil\relax}
  {\rightskip\dimexpr\tocleft@pagenum+1em\relax plus 1fil\relax}
  {}{\PackageError{layout}{Cannot adjust contents spacing}{}}
\makeatother
\begin{document}
\title{The Weyl double copy for relativistic Rindler fluids}
\author{Yu-Cheng Wang$^{1,3,4}$}
\email
{wangyucheng25@mails.ucas.ac.cn}
\author{Jing-Rui Zhang$^{1,3,4}$}
\email
{zhangjingrui22@mails.ucas.ac.cn}
\author{Yun-Long Zhang$^{2,1}$}
\email
{zhangyunlong@nao.cas.cn}
\affiliation{$^{1}$ School of Fundamental Physics and Mathematical Sciences,  Hangzhou   Institute for Advanced Study, UCAS, Hangzhou 310024, China. }
\affiliation{$^{2}$ National Astronomical Observatories, Chinese Academy of Sciences, Beijing, 100101, China}
\affiliation{$^{3}$CAS Key Laboratory of Theoretical Physics, Institute of Theoretical Physics, Chinese Academy of Sciences, Beijing 100190, China.} 
\affiliation{$^{4}$Taiji Laboratory for Gravitational Wave Universe(Beijing/Hangzhou), University of Chinese Academy of Sciences, Beijing 100149, China.}

\begin{abstract}
We study the Petrov classification and Weyl double copy of the four-dimensional bulk metric dual to a $2+1$-dimensional relativistic Rindler fluid at first order in the relativistic gradient expansion. Upon imposing the fluid equations, the leading Weyl curvature is determined by the fluid vorticity and shear. At this order, the bulk geometry is generically of Petrov type II, with a type D branch for shear-free flows with nonzero vorticity and a type N branch for irrotational flows with nonzero shear. For the type D branch, we construct a purely magnetic Maxwell single copy on a fixed Rindler background and distinguish exact source-free solutions with a constant magnetic field from local perturbative solutions valid through first order in gradients. For the type N branch, we obtain a family of algebraic Weyl double copies parametrized by $\beta$. We derive explicit field strengths and gauge potentials for $\beta=0$ and $\beta=1$, together with the local Cauchy-Riemann conditions imposed by Maxwell's equations. These choices encode the fluid shear differently: for $\beta=0$, the shear dependence resides entirely in the zeroth-copy scalar, whereas for $\beta=1$, it enters the Maxwell field directly. Finally, using the non-relativistic hydrodynamic expansion with explicit $\epsilon$ counting, we derive the non-relativistic limits of the type D single copy and both type N choices, and compare the resulting structures with existing non-relativistic constructions.

\medskip
\noindent\textbf{Keywords:} Rindler fluid; fluid/gravity correspondence; Petrov classification; Weyl double copy

\end{abstract}
\maketitle
\tableofcontents

\section{Introduction}

In the fluid/gravity correspondence, long-wavelength fluid dynamics can be encoded in a bulk geometry with one extra dimension~\cite{Bhattacharyya2008,Rangamani2009,Hubeny2012}. For perturbations of Rindler spacetime, Dirichlet boundary conditions are imposed on a timelike cutoff surface at a fixed radial position $r=r_c$, together with regularity at the future horizon~\cite{Bredberg2011}. Under these conditions, the constraint components of the vacuum Einstein equations reduce to the incompressible Navier-Stokes equations~\cite{ElingFouxonOz2009,ElingOz2010,LysovStrominger2011,Huang2011}. A non-relativistic map from solutions of these fluid equations to Ricci-flat bulk metrics was constructed in Ref.~\cite{Bredberg2012}. In four dimensions, the resulting metrics were shown to be algebraically special of Petrov type II at the relevant perturbative order. This construction was subsequently extended to a covariant relativistic gradient expansion, in which the equilibrium pressure and velocity are promoted to slowly varying boundary fields and the bulk Einstein equations are solved order by order~\cite{Compere2012,Eling2012}.

The Brown-York stress tensor on a timelike cutoff surface provides an effective fluid description of the bulk geometry. On the Rindler branch, the resulting Rindler fluid has vanishing equilibrium energy density and nonzero equilibrium pressure~\cite{Compere2012,Eling2012,Compere2011}. Its near-equilibrium dynamics, described by Rindler hydrodynamics, is encoded in the associated fluid-dual spacetime~\cite{Keeler2020}. In equilibrium, the corresponding fluid-dual metric reduces to that of the flat Rindler background. Away from equilibrium, gradients of the fluid variables require corrections to this metric so that the bulk Einstein equations remain satisfied.

In the color-kinematics formulation of the double copy, scattering amplitudes in the gravity theory can be constructed from gauge theory~\cite{Bern2008,Bern2010}. Classical extensions establish corresponding relations between solutions in these two kinds of theories~\cite{Monteiro2014,Luna2019,White2021,Godazgar2021,Easson2021,Easson2023}. The different formulations and their applications are reviewed in Ref.~\cite{Adamo2022Snowmass}. In the Weyl double copy, the curvature is expressed through a quadratic relation between the four-dimensional Weyl spinor and the Maxwell spinor~\cite{Luna2019},
\begin{equation}
  C_{ABCD}=\frac{1}{S}f_{(AB}f_{CD)},
  \label{eq:wdc-intro}
\end{equation}
where $f_{AB}$ is the anti-self-dual field-strength spinor of the Maxwell field and $S$ is the complex zeroth-copy scalar field. This formulation was systematically developed for vacuum Petrov type D spacetimes and extended to a class of type N wave spacetimes~\cite{Luna2019,Godazgar2021}. The multiplicities of the principal null directions are specified by the Petrov classification, which therefore constrains the spinorial factorization used to construct the Maxwell single copy~\cite{White2021}.

The Weyl double copy was applied to bulk metrics dual to non-relativistic Rindler fluids in Ref.~\cite{Keeler2020}. Constant-vorticity flows specialize the geometry to type D, with a single copy corresponding to a uniform magnetic field. Irrotational potential flows correspond to type N, with the dependence on the velocity potential encoded primarily in the zeroth-copy scalar. These results motivate a further analysis of the relation between fluid dynamics and the double copy~\cite{CheungMangan2020}.

In this paper, we study the Weyl double copy for a $(2+1)$-dimensional relativistic Rindler fluid through first order in the relativistic gradient expansion. Starting from the relativistic fluid-dual metric of Refs.~\cite{Compere2012,Eling2012}, we compute the five Weyl scalars of the four-dimensional bulk geometry and determine its Petrov classification at this order. The calculation retains the relativistic fluid pressure $\mathrm{p}(x)$ and unit timelike velocity $u_a(x)$, and shows how the fluid vorticity and shear control the leading curvature after the fluid equations are imposed. We identify the generic type II branch and its type D, N, and O specializations.

We construct Maxwell single copies for the type D and type N branches of this relativistic geometry. For type D, we obtain a purely magnetic single copy and distinguish exact source-free solutions with constant magnetic field $\calB$ from local solutions valid through first order in gradients. For type N, we obtain an algebraic family parametrized by $\beta$, derive explicit field strengths and gauge potentials for $\beta=0,1$, and determine the associated Maxwell compatibility conditions. We then derive the non-relativistic limits of these relativistic results using the non-relativistic hydrodynamic expansion in slow variables, with explicit $\epsilon$ counting, allowing comparison with the existing non-relativistic constructions~\cite{Keeler2020,KeelerMonga2025}.

The paper is organized as follows. Section~\ref{sec:geometry} presents the relativistic fluid-dual metric and the fluid constraints required by the vacuum Einstein equations. Section~\ref{sec:weyl} specifies the null tetrad and curvature conventions, computes the Weyl scalars, and determines the Petrov classification. Section~\ref{sec:wdc} sets out the spinor conventions and the algebraic and differential conditions for the Weyl double copy. Section~\ref{sec:typeD} constructs the type D single copy and analyzes its Maxwell equations, while Section~\ref{sec:typeN} develops the type N family and its explicit solutions. Both sections also derive the corresponding non-relativistic limits. Section~\ref{sec:conclusion} summarizes the results and discusses future directions. Appendices~\ref{app:seed}-\ref{app:NR-metric} provide details of the Rindler seed metric, the hydrodynamic frame and first-order metric correction, and the non-relativistic fluid-dual metric used for comparison.

\section{Relativistic Rindler fluid and dual metric}
\label{sec:geometry}

In this section, the coordinates, fluid kinematic quantities, and order counting in the relativistic gradient expansion are specified, and the seed metric and its first-order correction are presented. We use the convention $16\pi G=1$ throughout. The fluid constraints are derived from conservation of the Brown-York stress tensor, and their role in cancelling the first-order contribution to the Ricci tensor is exhibited in local rest-frame coordinates.

\subsection{Relativistic gradient expansion}

The four-dimensional bulk coordinates are denoted by
\begin{equation}
  x^\mu=(r,x^a)=(r,\tau,x,y),
  \label{eq:coordinates}
\end{equation}
where $x^a$ denotes the coordinates on the three-dimensional timelike cutoff surface. Here and below, boundary quantities are defined on this finite cutoff surface. The cutoff radius is set to $r_c=1$, so that the induced metric and velocity normalization are
\begin{equation}
  \eta_{ab}=\operatorname{diag}(-1,1,1),
  \qquad
  \eta^{ab}u_a u_b=-1.
\end{equation}
The transverse projector, convective derivative, transverse derivative, and fluid acceleration are defined by
\begin{equation}\begin{aligned}
  &h_{ab}=\eta_{ab}+u_a u_b,
  \quad
  D=u^a\partial_a,
  \\
  &D_a^\perp=h_a{}^b\partial_b,\qquad
  a_a=Du_a.
  \label{eq:kinematics-1}
\end{aligned}\end{equation}
The fluid shear tensor, expansion scalar, and vorticity tensor are defined by
\begin{equation}\begin{aligned}
  &K_{ab}=h_a{}^c h_b{}^d\partial_{(c}u_{d)},
  \quad
  \Theta=\partial_a u^a,
  \\
  &\Omega_{ab}=h_a{}^c h_b{}^d\partial_{[c}u_{d]}.
  \label{eq:kinematics-2}
\end{aligned}\end{equation}
 With the leading-order incompressibility condition imposed below, \(K_{ab}\) is traceless to the order retained and is referred to as the fluid shear tensor~\cite{Compere2011,Compere2012}. It is worth noting that \(K_{ab}\) does not denote the extrinsic curvature of the cutoff surface.

In the relativistic gradient expansion, the fluid pressure $\mathrm{p}$ and velocity $u_a$ are zeroth-order quantities, and each boundary derivative $\partial_a$ raises the order by one~\cite{Compere2012,Eling2012,Bhattacharyya2008,MeyerOz2013}. Radial derivatives do not raise the order in this expansion. Hence,
\begin{equation}
  K_{ab},\ \Omega_{ab},\ a_a,\ D\mathrm{p},\ D_a^\perp \mathrm{p}
  \sim\Order(\partial).
\end{equation}

\subsection{Seed metric and first-order correction}

After a radial shift, a time rescaling, and a boundary Lorentz boost(see Appendix~\ref{app:seed}), the boosted Rindler seed metric can be written as~\cite{Compere2012}
\begin{equation}
  \begin{aligned}
  \dd s_{(0)}^2
  ={}&-2\mathrm{p} u_a\dd x^a\dd r\\
  &+\left[\eta_{ab}-\mathrm{p}^2(r-1)u_a u_b\right]\dd x^a\dd x^b.
  \end{aligned}
  \label{eq:seed-metric}
\end{equation}
Introducing
\begin{equation}
  \tilde{r} (r)\equiv 1+\mathrm{p}^2(r-1),
  \label{eq:hatr}
\end{equation}
the seed metric can be written equivalently as
\begin{equation}
  \dd s_{(0)}^2
  =-2\mathrm{p} u_a\dd x^a\dd r
  +\left(h_{ab}-\tilde{r} u_a u_b\right)\dd x^a\dd x^b.
\end{equation}
The Rindler horizon is located at
\begin{equation}
  r_H=1-\frac{1}{\mathrm{p}^2}.
\end{equation}
For constant $\mathrm{p}$ and $u_a$, the seed metric describes flat spacetime in Rindler coordinates. Promoting these parameters to slowly varying fields produces a nonzero Ricci tensor at first order in gradients. In the radial gauge
\begin{equation}
  g_{rr}=0,
  \qquad
  g_{ra}=-\mathrm{p}u_a,
\end{equation}
the requirement that the induced metric at $r=1$ remains fixed leads to the first-order metric correction, denoted by $\delta g_{ab}$ (see Appendix~\ref{app:first-order})~\cite{Compere2012}
\begin{equation}
  \delta g_{ab}
  =2(r-1)\left[u_a u_b D\mathrm{p}+2\mathrm{p} a_{(a}u_{b)}\right].
  \label{eq:first-correction}
\end{equation}
The factor \(r-1\) ensures that the correction vanishes on the cutoff surface, leaving the prescribed induced metric unchanged. The remaining factors encode first-order departures from equilibrium through the pressure variation along the flow and the fluid acceleration.
The bulk metric dual to the relativistic Rindler fluid, through first order in gradients, is therefore
\begin{equation}
  {\begin{aligned}
  \dd s^2={}&-2\mathrm{p}u_a\dd x^a\dd r 
  +\Bigl[\eta_{ab}-\mathrm{p}^2(r-1)u_a u_b\\
  &+2(r-1)\bigl(u_a u_bD\mathrm{p}
  +2\mathrm{p}a_{(a}u_{b)}\bigr)\Bigr]\dd x^a\dd x^b\\
  &+\Order(\partial^2).
  \end{aligned}}
  \label{eq:rel-metric}
\end{equation}

\subsection{Rindler fluid and Einstein equations}

The fluid stress tensor is identified with the Brown-York stress tensor on the cutoff surface. With the fluid variables fixed as in Appendix~\ref{app:first-order}, its first-order constitutive relation is~\cite{Compere2012,BrownYork1993}
\begin{equation}
  T_{ab}=\mathrm{p} h_{ab}-2K_{ab}+\Order(\partial^2).
  \label{eq:BY}
\end{equation}
Conservation of its perfect-fluid (zeroth-order) part gives~\cite{Compere2012}
\begin{equation}\begin{aligned}
  &\Theta=\partial_a u^a\sim\Order(\partial^2),
  \\
  &\mathcal{E}_a\equiv a_a+D_a^\perp\ln \mathrm{p}\sim\Order(\partial^2).
  \label{eq:fluid-eq}
\end{aligned}\end{equation}
These are the leading-order relativistic fluid equations, including the incompressibility condition and the transverse Euler equation, respectively. The explicit dependence on \(\Theta\) and \(\mathcal{E}_a\) identifies the terms in the bulk vacuum Einstein equations that are eliminated by the fluid constraints.

At a selected boundary point $x_*^a$, local rest-frame coordinates
are introduced by a boundary Lorentz transformation with constant
parameters, leaving the radial coordinate $r$ unchanged.
For the boundary metric $\eta_{ab}=\operatorname{diag}(-1,1,1)$,
the velocity components at that point satisfy
\begin{equation}
  \begin{aligned}
    &u^a\big|_{x_*}=(1,0,0),\quad\
    u_a\big|_{x_*}=(-1,0,0),
  \end{aligned}
\end{equation}
where the components are ordered as $(\tau,x,y)$.
Equivalently,
\begin{equation}
  \left.u_a\,\mathrm{d}x^a\right|_{x_*}
  =-\mathrm{d}\tau.
\end{equation}
This condition fixes the velocity at the selected point without
requiring the velocity field to be constant in a neighborhood.

To define spatial projections, we introduce an orthonormal frame
$e_{I}=e_{I}{}^a\partial_a$ in the boundary spatial
plane orthogonal to $u^a$, with $I,J=1,2$:
\begin{equation}
\begin{aligned}
  u_a e_{I}{}^a&=0,\\
  \eta_{ab}e_{I}{}^a e_{J}{}^b&=\delta_{IJ}.
\end{aligned}
\label{eq:boundary-spatial-frame}
\end{equation}
Here $I=1,2$ labels the two boundary spatial
orthonormal-frame directions, whereas $a$ is a boundary
coordinate index. The corresponding spatial coframe is defined by
\begin{equation}
\begin{aligned}
  e^{I}&=e^{I}{}_a\,\dd x^a,\\
  e^{I}{}_a
  &=\delta^{IJ}\eta_{ab}e_{J}{}^b,
\end{aligned}
\label{eq:boundary-spatial-coframe}
\end{equation}
so that
\begin{equation}
  e^{I}(e_{J})=\delta^{I}{}_{J},
  \qquad
  e^{I}(u)=0.
\label{eq:boundary-frame-duality}
\end{equation}
The spatial projector can therefore be written as
$h_{ab}=\delta_{IJ}e^{I}{}_a e^{J}{}_b$.

The spatial components of the transverse Euler-equation
residual in Eq.~\eqref{eq:fluid-eq} are defined by
\begin{equation}
  \mathcal{E}_{I}
  \equiv \mathcal{E}_a e_{I}{}^a
  =\left(a_a+D_a^\perp\ln\mathrm{p}\right)e_{I}{}^a.
\label{eq:Euler-frame-components}
\end{equation}
The indices $I,J=1,2$ on projected boundary quantities refer to
these two spatial frame directions throughout.

At the selected point, the spatial frame is aligned with
the local rest-frame coordinates:
\begin{equation}
\begin{aligned}
  \left.e_{1}\right|_{x_*}&=\partial_x,
  &\left.e_{2}\right|_{x_*}&=\partial_y,\\
  \left.e^{1}\right|_{x_*}&=\dd x,
  &\left.e^{2}\right|_{x_*}&=\dd y.
\end{aligned}
\label{eq:boundary-frame-alignment}
\end{equation}
Thus, $\mathcal{E}_{1}=\mathcal{E}_x$ and
$\mathcal{E}_{2}=\mathcal{E}_y$ at that point.

Under these local rest-frame conditions, with bulk coordinates
ordered as $(r,\tau,x,y)$, the first-order contribution to the
Ricci tensor before imposing the fluid equations
\eqref{eq:fluid-eq} can be expressed as
\begin{equation}
  \left.R_{\mu\nu}\right|_{\Order(\partial)}
  =\frac{\mathrm{p}}{2}
  \begin{pmatrix}
  0&0&0&0\\
  0&\Theta&-\mathcal{E}_{1}&-\mathcal{E}_{2}\\
  0&-\mathcal{E}_{1}&0&0\\
  0&-\mathcal{E}_{2}&0&0
  \end{pmatrix}.
  \label{eq:ricci-residual}
\end{equation}
Thus, upon imposing the fluid equations in Eq.~\eqref{eq:fluid-eq}, we obtain~\cite{Compere2012},
\begin{equation}
  R_{\mu\nu}\sim\Order(\partial^2),
  \qquad
  R\sim\Order(\partial^2).
\end{equation}
The vacuum condition is imposed order by order: the Ricci tensor vanishes at zeroth and first order in gradients, while terms at second and higher orders are beyond the truncation considered here.

\section{Weyl scalars and Petrov classification}
\label{sec:weyl}

In this section, the null tetrad and curvature conventions are specified, and the five Weyl scalars are evaluated at first order in the relativistic gradient expansion. After the fluid equations are imposed, the curvature is expressed in terms of vorticity and shear. The type II, D, N, and O branches are identified, together with the perturbative order at which the classification applies. The Weyl quantities entering the double copy construction below are understood at first order in gradients.

\subsection{Null tetrad and sign conventions}

Null-tetrad methods also underlie the Newman-Penrose map. Here, the curvature of the bulk geometry dual to the Rindler fluid is resolved into Weyl scalars using a null tetrad.

At each boundary point, a local rest frame is chosen such that
\begin{equation}
  u_a\dd x^a=-\dd\tau
\end{equation}
holds at that point. The zeroth-order seed metric then becomes
\begin{equation}
  \dd s_{(0)}^2=-\tilde{r}(r)\dd\tau^2+2\mathrm{p}\dd\tau\dd r+\dd x^2+\dd y^2.
  \label{eq:local-seed}
\end{equation}
A null coframe for the seed metric is given by
\begin{equation}
  \begin{aligned}
  \ell&=-\sqrt{\frac{\tilde{r}}{2}}\,\dd\tau,~~
  n=-\sqrt{\frac{\tilde{r}}{2}}\,\dd\tau
      +\mathrm{p}\sqrt{\frac{2}{\tilde{r}}}\,\dd r,
  \label{eq:null-tetrad-ln}\\
  m&=\frac{-\ii\,\dd x+\dd y}{\sqrt2},
  \qquad
  \bar m=m^*.
\end{aligned}
\end{equation}
It satisfies
\begin{equation}
  g^{(0)}=-\ell\otimes n-n\otimes\ell
  +m\otimes\bar m+\bar m\otimes m.
\end{equation}
The boundary spatial frame $e_{I}$ and its dual
coframe $e^{I}$ were defined in
Section~\ref{sec:geometry}.
The labels $I,J=1,2$ on quantities such as $K_{IJ}$ and
$\mathcal{E}_{I}$ refer to this boundary spatial frame.
The four-dimensional coordinates follow the ordering
$x^\mu=(r,x^a)=(r,\tau,x,y)$ in Eq.~\eqref{eq:coordinates}.
In the non-relativistic limit,
\begin{equation}
  \mathrm{p}\rightarrow1,
  \qquad
  \tilde{r}\rightarrow r.
\end{equation}
The null coframe in Eq.~\eqref{eq:null-tetrad-ln} coincides with that used in the Weyl double copy construction for non-relativistic Rindler fluids~\cite{Keeler2020}.

For the metric signature $(-,+,+,+)$,
the Newman-Penrose Weyl scalars are defined by~\cite{NewmanPenrose1962}
\begin{align}
 \Psi_0&=C_{\mu\nu\rho\sigma}\ell^\mu m^\nu\ell^\rho m^\sigma,
 \notag
 \Psi_1=C_{\mu\nu\rho\sigma}\ell^\mu n^\nu\ell^\rho m^\sigma,
 \notag\\
 \Psi_2&=C_{\mu\nu\rho\sigma}\ell^\mu m^\nu\bar m^\rho n^\sigma,
 \notag
 \Psi_3=C_{\mu\nu\rho\sigma}\ell^\mu n^\nu\bar m^\rho n^\sigma,
 \notag\\
 \Psi_4&=C_{\mu\nu\rho\sigma}n^\mu\bar m^\nu n^\rho\bar m^\sigma.
 \label{eq:NP-definitions}
\end{align}

\subsection{Weyl scalars at first order in gradients}

Using the boundary spatial frame defined above, the
projected velocity gradient and vorticity component are
\begin{equation}
\begin{aligned}
  &K_{IJ}=K_{ab}e_{I}{}^a e_{J}{}^b,
  \qquad I,J=1,2,\\
  &\Omega_{12}
  =\Omega_{ab}e_{1}{}^a e_{2}{}^b.
\end{aligned}
\end{equation}
Contraction of the Weyl tensor with the null tetrad gives, at first order in gradients,
\begin{align}
\Psi_0&\sim\Order(\partial^2),\qquad
  \Psi_1\sim\Order(\partial^2),
  \label{eq:psi01}\\
  \Psi_2&=\frac{\ii \mathrm{p}}{2}\Omega_{12}+\Order(\partial^2),
  \label{eq:psi2}\\
  \Psi_3&=-\frac{\ii \mathrm{p}}{4\sqrt{\tilde{r}}}
  \left(\mathcal{E}_{1}-\ii \mathcal{E}_{2}\right)+\Order(\partial^2),
  \label{eq:psi3}\\
  \Psi_4&=-\frac{\mathrm{p}}{2\tilde{r}}\left(K_{11} - K_{22} - 2\ii K_{12}\right) + \mathcal{O}(\partial^2).
  \label{eq:psi4}
\end{align}
These expressions show that distinct fluid kinematic quantities enter different Weyl components. The vorticity determines \(\Psi_2\), while the relevant combination of the symmetric velocity gradient determines \(\Psi_4\). The first-order contribution to \(\Psi_3\) is proportional to the residual \(\mathcal{E}_a\) of the transverse Euler equation in Eq.~\eqref{eq:fluid-eq} and therefore vanishes when the fluid equations are imposed.
The trace of the projected symmetric velocity gradient satisfies
\begin{equation}
\begin{aligned}
K_{11}+K_{22} &= h^{ab}K_{ab}\\
&= \partial_a u^a + \frac{1}{2} u^a \partial_a(u^b u_b) \\
&= \Theta \sim \Order(\partial^2).
\end{aligned}
\end{equation}
The final order estimate follows from the fluid constraints in Eq.~\eqref{eq:fluid-eq}. With these constraints imposed, the Weyl scalars reduce to
\begin{equation}\begin{aligned}
  &\Psi_0\sim \Order(\partial^2),\qquad\Psi_1\sim \Order(\partial^2),\\&\Psi_3\sim \Order(\partial^2),\qquad
  \Psi_2=\frac{\ii \mathrm{p}}{2}\Omega_{12}+\Order(\partial^2),
  \\
  &\Psi_4=-\frac{\mathrm{p}}{\tilde{r}}(K_{11}-\ii K_{12})+\Order(\partial^2).
  \label{eq:on-shell-psis}
\end{aligned}\end{equation}

\subsection{Petrov classification}

The resulting branches are summarized in Table~\ref{tab:petrov}. Here the Petrov classification is applied to the leading nonvanishing Weyl curvature at first order in the relativistic gradient expansion. Higher-order terms may lift the degeneracies of the principal null directions~\cite{Keeler2020,Cai2014,KeelerMonga2025,CaiLiYangZhang2013}.

\begin{table*}[t]
  \centering
  \caption{Petrov classification of the bulk geometry dual to a relativistic Rindler fluid at first order in the relativistic gradient expansion.}
  \label{tab:petrov}
  \begin{tabular}{ccc}
    \toprule
    Kinematic conditions & Nonzero Weyl scalars (leading order) & Petrov type \\
    \midrule
    $\Omega_{12}\neq0$, $K_{IJ}\neq0$ & $\Psi_2,\Psi_4$ & II \\
    $\Omega_{12}\neq0$, $K_{IJ}=0$ & $\Psi_2$ & D \\
    $\Omega_{12}=0$, $K_{IJ}\neq0$ & $\Psi_4$ & N \\
    $\Omega_{12}=0$, $K_{IJ}=0$ & No first-order Weyl curvature & O  \\
    \bottomrule
  \end{tabular}
\end{table*}

\section{Weyl double copy conventions}
\label{sec:wdc}

In this section, the Weyl and Maxwell spinors are expanded in a spin frame, and the algebraic relations between their scalar components are derived. Constant rescalings and electromagnetic duality rotations are discussed. The differential conditions imposed by Maxwell's equations and the zeroth-copy scalar wave equation are stated separately from the algebraic double copy relation.

The Weyl spinor is expanded in the spin frame $(o_A,\iota_A)$ as~\cite{NewmanPenrose1962,Keeler2020}
\begin{align}
 C_{ABCD}={}&\Psi_0\iota_A\iota_B\iota_C\iota_D
 -4\Psi_1o_{(A}\iota_B\iota_C\iota_{D)}
 \notag\\ &+6\Psi_2o_{(A}o_B\iota_C\iota_{D)}
 -4\Psi_3o_{(A}o_Bo_C\iota_{D)}\notag\\
 &+\Psi_4o_Ao_Bo_Co_D.
 \label{eq:weyl-spinor}
\end{align}
The Maxwell spinor is expanded in the same spin frame, with Newman-Penrose Maxwell scalars $\phi_0,\phi_1,\phi_2$, as
\begin{equation}
  f_{AB}=\phi_0\iota_A\iota_B
  -2\phi_1o_{(A}\iota_{B)}
  +\phi_2o_Ao_B.
  \label{eq:maxwell-spinor}
\end{equation}
Comparison of the spinor components of the Weyl double copy relation in Eq.~\eqref{eq:wdc-intro} gives
\begin{equation}
\begin{array}{c|c}
\text{Weyl scalar}&\text{Relation to Maxwell scalars}\\ \hline
\Psi_0&\phi_0^2/S\\
\Psi_1&\phi_0\phi_1/S\\
\Psi_2&(\phi_0\phi_2+2\phi_1^2)/(3S)\\
\Psi_3&\phi_1\phi_2/S\\
\Psi_4&\phi_2^2/S
\end{array}
\label{eq:scalar-wdc}
\end{equation}

The Weyl double copy relation in Eq.~\eqref{eq:wdc-intro} is invariant under the constant rescaling
\begin{equation}
  f_{AB}\rightarrow\lambda f_{AB},
  \qquad
  S\rightarrow\lambda^2S,
  \label{eq:wdc-scaling}
\end{equation}
and the constant electromagnetic duality rotation
\begin{equation}
  (\phi_0,\phi_1,\phi_2)
\rightarrow
e^{\mathrm i\theta}(\phi_0,\phi_1,\phi_2),\qquad S\rightarrow e^{2\mathrm i\theta}S.
  \label{eq:duality-phase}
\end{equation}
Both transformations leave the Weyl spinor unchanged.

The spinorial structure of the Weyl double copy also admits a twistor-space interpretation. At the level of linearized fields, the Penrose transform provides a framework for deriving spacetime double copy relations and studying their Petrov structure~\cite{White2021,Chacon2021}.

The scalar relations in Eq.~\eqref{eq:scalar-wdc} are algebraic. A complete Weyl double copy also requires the Maxwell equations and the zeroth-copy scalar wave equation to be satisfied~\cite{Luna2019,Godazgar2021,Keeler2020}. Fields satisfying the algebraic relation are constructed below, and the following differential conditions are evaluated on the fixed Rindler background at the relevant perturbative order:
\begin{equation}\begin{aligned}
  &\dd F=0,
  \
  \nabla_\mu F^{\mu\nu}=0,
  \
  \Box S=0.
  \label{eq:field-equations}
\end{aligned}\end{equation}

\section{Petrov type D branch}
\label{sec:typeD}

In this section, a purely magnetic Maxwell single copy and its zeroth-copy scalar are constructed from \(\calB=\mathrm{p}\Omega_{12}\) for the branch with vanishing shear and nonzero vorticity. Exact source-free solutions with constant \(\calB\) are distinguished from local solutions valid to first order in gradients with slowly varying \(\calB\). The non-relativistic limit is derived, and the uniform magnetic field associated with constant vorticity is identified.

\subsection{Type D conditions and double copy}

The type D branch satisfies~\cite{Keeler2020}
\begin{equation}
  K_{IJ}=0,
  \qquad
  \Omega_{12}\neq0.
\end{equation}
The vorticity-dependent scalar is defined by
\begin{equation}
  \calB(x^a)\equiv \mathrm{p}(x^a)\Omega_{12}(x^a).
  \label{eq:q-def}
\end{equation}
Since $\Omega_{12}\sim\Order(\partial)$,
\begin{equation}
  \calB\sim\Order(\partial),
  \qquad
  \partial_a \calB\sim\Order(\partial^2).
  \label{eq:q-order}
\end{equation}
The scalar \(\calB\) contributes to the algebraic construction at first order in gradients, whereas its boundary derivatives enter Maxwell's equations at second order in gradients. Neglecting those derivatives in a first-order truncation therefore imposes no exact constancy condition on \(\calB\). The nonvanishing Weyl scalar and the Weyl spinor entering the double copy relation in Eq.~\eqref{eq:wdc-intro} are
\begin{equation}
  \Psi_2=\frac{\ii}{2}\calB,
  \quad
  C_{ABCD}=3\ii \calB\,o_{(A}o_B\iota_C\iota_{D)}.
\end{equation}
A Maxwell spinor satisfying the algebraic double copy relation can be chosen with
\begin{equation}
  \phi_0=\phi_2=0,
  \qquad
  \phi_1=-\frac{\ii c_Be^{\ii\theta}}{2}\calB,
\end{equation}
for which $\Psi_2=2\phi_1^2/(3S)$ fixes the zeroth-copy scalar as
\begin{equation}
  S^{(D)}=\frac{\ii c_B^2e^{2\ii\theta}}{3}\calB.
  \label{eq:SD}
\end{equation}
Here, $c_B$ is a nonzero real normalization constant and $\theta$ is a constant electromagnetic duality angle. The two principal null directions of the Maxwell field therefore coincide with the two repeated principal null directions of the type D Weyl tensor. With the convention
\begin{equation}
  \phi_1=\frac12(E_{r}-\ii B_{r}),
\end{equation}
the choice $\theta=0$ yields a purely magnetic solution: $E_{r}=0$ and $B_{r}=c_B\calB$.

Electric and magnetic sectors also arise in the classical double copy of Taub-NUT spacetime, whose dyonic single copy relates electric and magnetic charges to the gravitational mass and NUT charge~\cite{Luna2015TaubNUT}.

\subsection{Maxwell single copy and field equations}

For the purely magnetic phase, the boundary spatial
coframe is held fixed at the chosen local rest-frame
point and extended independently of $r$ onto the fixed
Rindler background. Its extensions are denoted by the
same symbols, with $e^{1}=\dd x$ and
$e^{2}=\dd y$. The field strength is then~\cite{Keeler2020}
\begin{equation}
  F^{(D)}=c_B\calB(x^a)e^{1}\wedge e^{2}.
\end{equation}
A general radial extension of \(\mathcal B\) is considered, and Maxwell's equations require it to be independent of the radial coordinate. In ingoing Rindler coordinates, with the orientation adopted here, this becomes
\begin{equation}
  F^{(D)}=c_B\calB(r,\tau,x,y)\,\dd x\wedge\dd y.
  \label{eq:FD-general}
\end{equation}
The homogeneous Maxwell equation gives
\begin{align}
 \dd F^{(D)}
 ={}&c_B(\partial_\tau \calB)\dd\tau\wedge\dd x\wedge\dd y\notag\\
 &+c_B(\partial_r \calB)\dd r\wedge\dd x\wedge\dd y,
\end{align}
and hence requires
\begin{equation}
  \partial_\tau \calB=\partial_r\calB=0.
\end{equation}
The Maxwell single copy is defined on the fixed Rindler background, obtained by holding the zeroth-order seed data constant and omitting the fluid perturbations~\cite{Keeler2020,KeelerMonga2025}. Thus, $\mathrm{p}$ and the local frame data are held fixed in the Maxwell operator. The fluid vorticity enters the field strength only through $\calB$. The full velocity field that generates this vorticity is not additionally incorporated into the fixed Rindler background. The distinction between the full gravitational geometry and the background used for its single copy is also explicit in Kerr-Schild constructions on curved backgrounds~\cite{BahjatAbbas2017}. In the present calculation, the fixed Rindler background is flat. On this background, the nontrivial components of the inhomogeneous Maxwell equation are
\begin{equation}
  \nabla_\mu^{(0)}F_{(D)}^{\mu x}=-c_B\partial_y\calB,
  \qquad
  \nabla_\mu^{(0)}F_{(D)}^{\mu y}=c_B\partial_x\calB.
\end{equation}
Consequently, an exact source-free solution with only an $F_{xy}$ component requires
\begin{equation}
  \calB=\calB_0=\text{constant}.
\end{equation}
A corresponding gauge potential is
\begin{equation}
  A^{(D)}=\frac{c_B\calB_0}{2}(x\,\dd y-y\,\dd x)+\dd\mathcal{C},
  \label{eq:AD}
\end{equation}
where \(\mathcal{C}(x^\mu)\) is an arbitrary smooth real gauge function and \(\dd\mathcal{C}=(\partial_\mu\mathcal{C})\dd x^\mu\) is its exterior derivative. This exact one-form represents the gauge freedom of the potential and leaves \(F{}=\dd A{}\) unchanged because \(\dd^2\mathcal{C}=0\). The corresponding field strength and zeroth-copy scalar are
\begin{equation}\begin{aligned}
  &F^{(D)}=\dd A^{(D)}=c_B\calB_0\dd x\wedge\dd y,
  \\
  &S^{(D)}=\frac{\ii c_B^2\calB_0}{3},
  \quad
  \Box S^{(D)}=0.
\end{aligned}\end{equation}

If $\calB$ is a slowly varying first-order field satisfying Eq.~\eqref{eq:q-order}, or if the background data $\mathrm{p},u_a$ themselves vary slowly, the residuals in Maxwell's equations arising from $\partial \calB$, $\calB\,\partial \mathrm{p}$, and variations of the frame begin at $\Order(\partial^2)$. The field therefore still defines a consistent local perturbative single copy at the first-order accuracy considered here, but it is no longer an exact source-free Maxwell solution valid to all orders. Retaining these variations at second and higher orders requires solving the higher-order Maxwell and scalar equations and reexamining the Petrov type D conditions. Additional components of $F_{\mu\nu}$ or a nonzero current may then be necessary.

\subsection{non-relativistic limit of the single copy}
\label{sec:typeD-NR-limit}

The non-relativistic variables used for both the type D and type N branches are introduced first. The non-relativistic hydrodynamic expansion is organized by the scaling~\cite{Bredberg2012,Keeler2020,Bhattacharyya2009}
\begin{equation}\begin{aligned}
  &\partial_i\sim\epsilon,
  \
  \partial_\tau\sim\epsilon^2,
  \
  u_i\sim\epsilon,
  \
  \mathrm{p}-1\sim\epsilon^2.
  \label{eq:NR-scaling}
\end{aligned}\end{equation}
The slow variables are defined by
\begin{equation}\begin{aligned}
  &T=\epsilon^2\tau,
  \
  X=\epsilon x,
  \
  Y=\epsilon y,
  \
  Z=X+\ii Y.
  \label{eq:slow-coordinates}
\end{aligned}\end{equation}
These slow variables make the long-wavelength and slow-time dependence explicit. By the chain rule, a spatial derivative acting on a function of the slow variables contributes one power of \(\epsilon\), whereas a time derivative contributes two. The order of each quantity in the non-relativistic hydrodynamic expansion can therefore be determined directly from its velocity factors and derivative structure.
With $i=1,2$ denoting spatial coordinate components in the
$(x,y)$ and $(X,Y)$ systems, the relativistic velocity and pressure expand as
\begin{equation}\begin{aligned}
  &u_i^{\rel}(\tau,x,y)
  =\epsilon v_i(T,X,Y)+\Order(\epsilon^3),
  \\
  &\mathrm{p}=1+\epsilon^2P(T,X,Y)+\Order(\epsilon^4).
  \label{eq:u-p-NR}
\end{aligned}\end{equation}
Here $P$ is the non-relativistic pressure fluctuation about the equilibrium value $\mathrm{p}=1$, expressed in the slow variables~\cite{Compere2011,Compere2012}. Both $v_i$ and $P$ are held at order unity as $\epsilon\to0$.

For the type D branch, the velocity and pressure expansions in Eq.~\eqref{eq:u-p-NR}, together with the slow coordinates in Eq.~\eqref{eq:slow-coordinates}, give
\begin{equation}
  \calB = \mathrm{p}\Omega_{12}^{\rel}
  =\frac{\epsilon^2}{2}
  (\partial_Xv_Y-\partial_Yv_X)+\Order(\epsilon^4).
  \label{eq:B-NR-scaling}
\end{equation}
Consistently with the first-order counting in Eq.~\eqref{eq:q-order}, substituting this expression into \(\Psi_2=\ii\calB/2\) gives
\begin{equation}
  \Psi_2^{\rel}
  =\frac{\ii\epsilon^2}{4}
  (\partial_Xv_Y-\partial_Yv_X)+\Order(\epsilon^3).
  \label{eq:psi2-NR}
\end{equation}
With the vorticity convention of Ref.~\cite{Keeler2020},
\begin{equation}
  \partial_X v_Y - \partial_Y v_X \equiv 2\omega,
  \label{eq:NR-vorticity}
\end{equation}
the Weyl scalar takes the form
\begin{equation}
\Psi_2^{\rel} = \ii \epsilon^2 \frac{\omega}{2} + \mathcal{O}(\epsilon^3).
\end{equation}

After the overall factor of $\epsilon^2$ is extracted, the constant-vorticity type D solution corresponds to a uniform magnetic field normal to the fluid plane. This agrees with the single copy of the bulk metric dual to the non-relativistic fluid~\cite{Keeler2020}.

\section{Petrov type N branch}
\label{sec:typeN}

In this section, a family of algebraic double copies parametrized by \(\beta\) is constructed from the complex shear, and the conditions imposed by Maxwell's equations are derived. The field strengths and gauge potentials for \(\beta=1\) and \(\beta=0\) are obtained explicitly. The dependence of the Maxwell field and the zeroth-copy scalar on the fluid shear is compared between these cases, and their non-relativistic limits are derived using the non-relativistic hydrodynamic expansion in slow variables, with explicit \(\epsilon\) counting.

\subsection{Type N conditions and double copy}

The irrotational type N branch satisfies~\cite{Keeler2020}
\begin{equation}
  \Omega_{12}=0,
  \qquad
  K_{IJ}\neq0.
  \label{eq:typeN-conditions}
\end{equation}
The complex shear combination and the nonvanishing Weyl scalar are related by
\begin{equation}
  \Kc\equiv K_{11}-\ii K_{12},
  \qquad
  \Psi_4=-\frac{\mathrm{p}}{\tilde{r}}\Kc.
  \label{eq:complex-shear}
\end{equation}
The Weyl spinor entering the double copy relation in Eq.~\eqref{eq:wdc-intro} then reduces to
\begin{equation}
  C_{ABCD}=\Psi_4\,o_Ao_Bo_Co_D,
\end{equation}
so that all four principal spinors coincide. The scalar double copy relations then require~\cite{Godazgar2021,Keeler2020}
\begin{equation}\begin{aligned}
  &\phi_0=\phi_1=0,
  \quad
  \phi_2\neq0,
  \quad
  \frac{\phi_2^2}{S}=\Psi_4.
  \label{eq:typeN-algebraic}
\end{aligned}\end{equation}

\subsection{Maxwell single copy and field equations}

In a local region where $\Kc\neq0$, a family of algebraic double copies can be defined as
\begin{align}
  \phi_2^{(N_{\beta})}
  &=\frac{c_{\beta} e^{\ii\theta}}{\sqrt{\tilde{r}}}\Kc^{\beta},
  \label{eq:phi2-family}\\
  S^{(N_{\beta})}
  &=-\frac{c_{\beta}^2e^{2\ii\theta}}{\mathrm{p}}\Kc^{2\beta-1}.
  \label{eq:S-family}
\end{align}
The parameter \(\beta\) controls how the complex shear is distributed between the Maxwell spinor and the zeroth-copy scalar. Their individual dependence on \(\mathcal K\) varies with \(\beta\), but the powers in the double copy ratio always combine as \(2\beta-(2\beta-1)=1\). Consequently, this family satisfies the following Weyl double copy relation:
\begin{equation}
  \frac{\bigl(\phi_2^{(N_{\beta})}\bigr)^2}{S^{(N_{\beta})}}
  =-\frac{\mathrm{p}}{\tilde{r}}\Kc
  =\Psi_4.
\end{equation}
Thus, the algebraic relation holds for arbitrary $\beta$. For noninteger or negative integer $\beta$, $\phi_2^{(N_{\beta})}$ and $S^{(N_{\beta})}$ may exhibit divergences, poles, or branch ambiguities. The analysis below is restricted to $\beta=0$ and $\beta=1$, with the non-relativistic limit evaluated for $\beta=0$.

The nonzero field-strength components are~\cite{KeelerMonga2025}
\begin{equation}\begin{aligned}
  F^{(N_{\beta})}_{02}=F^{(N_{\beta})}_{12}&=-\operatorname{Im}\phi_2^{(N_{\beta})},\\
  F^{(N_{\beta})}_{03}=F^{(N_{\beta})}_{13}&=\operatorname{Re}\phi_2^{(N_{\beta})}.
\end{aligned}
\end{equation}
Substitution gives
\begin{equation}
  \begin{aligned}
  F^{(N_{\beta})}={}&-{\sqrt{\tilde{r}}}\dd\tau\wedge
  \Bigl[\operatorname{Im}\phi_2^{(N_{\beta})}\dd x
+\operatorname{Re}\phi_2^{(N_{\beta})}\dd y\Bigr].
  \end{aligned}
  \label{eq:F-from-phi2}
\end{equation}
By Eq.~\eqref{eq:phi2-family}, $\sqrt{\tilde{r}}\,\phi_2^{(N_{\beta})}=c_{\beta}e^{\ii\theta}\Kc^{\beta}$, so the radial dependence cancels from the field strength.
On the fixed Rindler background, with $\mathrm{p}$ held constant, the equations $\dd F=0$ and
$\nabla_\mu F^{\mu\nu}=0$ respectively impose
\begin{equation}
  \begin{aligned}
    \partial_r\!\left(\sqrt{\tilde{r}}\,\operatorname{Re}\phi_2^{(N_{\beta})}\right)&=0,\\
    \partial_r\!\left(\sqrt{\tilde{r}}\,\operatorname{Im}\phi_2^{(N_{\beta})}\right)&=0,\\
    \partial_x\operatorname{Re}\phi_2^{(N_{\beta})}
      -\partial_y\operatorname{Im}\phi_2^{(N_{\beta})}&=0,\\
    \partial_y\operatorname{Re}\phi_2^{(N_{\beta})}
      +\partial_x\operatorname{Im}\phi_2^{(N_{\beta})}&=0.
  \end{aligned}
  \label{eq:CR}
\end{equation}
The last two equations in Eq.~\eqref{eq:CR} are the Cauchy-Riemann conditions for the complex Maxwell amplitude in the transverse plane. The radial conditions are automatically satisfied by Eq.~\eqref{eq:phi2-family}.

\subsection{\texorpdfstring{Type N solution with $\beta=1$}{Type N solution with beta=1}}

Setting $\beta=1$ gives
\begin{align}
  \phi_2^{(N_1)}
  ={}&\frac{c_1e^{\ii\theta}}{\sqrt{\tilde{r}}}\Kc 
  =\frac{c_1}{\sqrt{\tilde{r}}}
\Bigl[(K_{11}\cos\theta+K_{12}\sin\theta)\notag\\
  &\qquad\qquad +\ii(K_{11}\sin\theta-K_{12}\cos\theta)\Bigr],
  \label{eq:phi2-a1}\\
  S^{(N_1)}
  ={}&-\frac{c_1^2e^{2\ii\theta}}{\mathrm{p}}\Kc.
  \label{eq:S-a1}
\end{align}
Therefore,
\begin{align}
  \operatorname{Re}\phi_2^{(N_1)}
  &=\frac{c_1}{\sqrt{\tilde{r}}}
  (K_{11}\cos\theta+K_{12}\sin\theta),\\
  \operatorname{Im}\phi_2^{(N_1)}
  &=\frac{c_1}{\sqrt{\tilde{r}}}
  (K_{11}\sin\theta-K_{12}\cos\theta).
\end{align}
The corresponding field-strength two-form is
\begin{equation}
  {\begin{aligned}
  F^{(N_1)}={}&-{c_1}\dd\tau\wedge
  \Bigl[(K_{11}\sin\theta-K_{12}\cos\theta)\dd x\\
  &+(K_{11}\cos\theta+K_{12}\sin\theta)\dd y\Bigr].
  \end{aligned}}
  \label{eq:F-a1}
\end{equation}

In local rest-frame coordinates, the irrotational condition in Eq.~\eqref{eq:typeN-conditions} and the vanishing expansion required by the fluid equations~\eqref{eq:fluid-eq} imply, at the order considered,
\begin{equation}\begin{aligned}
  &K_{11}=\partial_xu_x,
  \
  K_{12}=\partial_xu_y=\partial_yu_x,
  \
  \partial_yu_y=-K_{11}.
\end{aligned}\end{equation}
An auxiliary scalar is defined by
\begin{equation}
  N_1=c_1(\sin\theta\,u_x-\cos\theta\,u_y).
\end{equation}
Then
\begin{align}
  \partial_xN_1=c_1(K_{11}\sin\theta-K_{12}\cos\theta),\\
\partial_yN_1=c_1(K_{11}\cos\theta+K_{12}\sin\theta).
\end{align}
The field strength can therefore be written as
\begin{equation}
  F^{(N_1)}=-\dd\tau\wedge\dd N_1.
\end{equation}
A corresponding gauge potential is
\begin{equation}
  {A^{(N_1)}={c_1}
  (\sin\theta\,u_x-\cos\theta\,u_y)\dd\tau+\dd\mathcal{C}.}
  \label{eq:A-a1}
\end{equation}
Thus, $\dd A^{(N_1)}=F^{(N_1)}$.

\subsection{\texorpdfstring{Type N solution with $\beta=0$}{Type N solution with beta=0}}

Setting $\beta=0$ gives~\cite{Keeler2020}
\begin{equation}
  \phi_2^{(N_0)}=\frac{c_0e^{\ii\theta}}{\sqrt{\tilde{r}}},
  \qquad
  S^{(N_0)}=-\frac{c_0^2e^{2\ii\theta}}{\mathrm{p}\Kc}.
  \label{eq:a0-phi-S}
\end{equation}
The associated field-strength two-form is
\begin{equation}
  {F^{(N_0)}=-{c_0}\dd\tau\wedge
  (\sin\theta\,\dd x+\cos\theta\,\dd y).}
  \label{eq:F-a0}
\end{equation}
A gauge potential is
\begin{equation}
  {A^{(N_0)}={c_0}
  (x\sin\theta+y\cos\theta)\dd\tau+\dd\mathcal{C}.}
  \label{eq:A-a0}
\end{equation}
The potential satisfies $\dd A^{(N_0)}=F^{(N_0)}$, and the resulting two-form has constant coefficients and satisfies the source-free Maxwell equations on the fixed Rindler background. For $\beta=0$, the Maxwell field is independent of $\Kc$,and the dependence on the velocity potential is contained
in the zeroth-copy scalar given in Eq.~\eqref{eq:a0-phi-S}.
This agrees with the structure of the non-relativistic potential-flow single copy~\cite{Keeler2020}.

\subsection{Non-relativistic limit of the single copy}
\label{sec:NR-limit}

The non-relativistic limits of both $\beta=1$ and $\beta=0$ are now derived using the scaling, slow coordinates, and fluid expansions in Eqs.~\eqref{eq:NR-scaling}-\eqref{eq:u-p-NR}.

For the type N branch, the non-relativistic flow is incompressible and irrotational, with nonzero shear as specified in Eq.~\eqref{eq:typeN-conditions}. Locally, its velocity can therefore be written in terms of a real velocity potential $\Phi(T,X,Y)$~\cite{Keeler2020}:
\begin{equation}\begin{aligned}
  &v_i=\partial_{X^i}\Phi,
  \\
  &\partial_Xv_Y-\partial_Yv_X=0,
  \
  \partial_Xv_X+\partial_Yv_Y=0.
  \label{eq:NR-potential-flow}
\end{aligned}\end{equation}
Here $X^i=(X,Y)$, so $\partial_{X^i}$ denotes differentiation with respect to $X$ or $Y$. Incompressibility then requires a harmonic potential, which can locally be expressed as
\begin{equation}\begin{aligned}
  &(\partial_X^2+\partial_Y^2)\Phi=0,
  \\
  &\Phi=f(T,Z)+\bar f(T,\bar Z).
  \label{eq:harmonic-potential}
\end{aligned}\end{equation}
At each fixed $T$, $f$ is holomorphic in $Z$. The complex derivatives are defined by
\begin{equation}
  \partial_Z=\frac12(\partial_X-\ii\partial_Y),
  \qquad
  \partial_{\bar Z}=\frac12(\partial_X+\ii\partial_Y),
  \label{eq:NR-complex-derivatives}
\end{equation}
which, together with the harmonicity of $\Phi$, give
\begin{align}
  2\partial_Z^2\Phi
  &=\partial_X^2\Phi-\ii\partial_X\partial_Y\Phi,
  \notag\\
  2\partial_{\bar Z}^2\Phi
  &=\partial_X^2\Phi+\ii\partial_X\partial_Y\Phi.
  \label{eq:complex-derivative}
\end{align}
Combining the potential-flow relations and complex derivatives in Eqs.~\eqref{eq:NR-potential-flow}-\eqref{eq:complex-derivative} with the complex shear defined in Eq.~\eqref{eq:complex-shear}, and using $\partial_x=\epsilon\partial_X$ and $\partial_y=\epsilon\partial_Y$, gives
\begin{align}
  \mathcal{K}^{\rel}
  &=K_{11}^{\rel}-\ii K_{12}^{\rel}\notag\\
  &=\epsilon^2
  \left(\partial_X^2\Phi-\ii\partial_X\partial_Y\Phi\right)
  +\Order(\epsilon^4)\notag\\
  &=2\epsilon^2\partial_Z^2f(T,Z)
  +\Order(\epsilon^4).
  \label{eq:Km-NR-scaling}
\end{align}
The coefficient of the leading non-relativistic shear is defined by
\begin{equation}
  \KNR
  \equiv\lim_{\epsilon\to0}\epsilon^{-2}\Kc^{\rel}.
\end{equation}
Then
\begin{equation}
  {\KNR=2\partial_{Z}^2 f(T,Z).}
  \label{eq:KNR}
\end{equation}

The definition of $\tilde{r}$ in Eq.~\eqref{eq:hatr} and the pressure expansion in Eq.~\eqref{eq:u-p-NR} give
\begin{equation}
  \tilde{r}=r+2\epsilon^2P(r-1)+\Order(\epsilon^4),
  \label{eq:hatr-NR}
\end{equation}
Substituting these expansions into Eq.~\eqref{eq:complex-shear} gives
\begin{align}
  \Psi_4^{\rel}
  &=-\frac{\mathrm{p}}{\tilde{r}}\mathcal{K}^{\rel}
  =-\frac{2\epsilon^2}{r}
  \partial_Z^2f(T,Z)
  +\Order(\epsilon^3).
  \label{eq:psi4-NR-limit}
\end{align}
The leading non-relativistic Weyl scalar is defined by
\begin{equation}
  \Psi_{4,\NR}\equiv
  \lim_{\epsilon\to0}\epsilon^{-2}\Psi_4^{\rel},
\end{equation}
and therefore satisfies
\begin{equation}
  {\Psi_{4,\NR}
  =-\frac{2}{r}\partial_Z^2f(T,Z).}
  \label{eq:psi4-NR-coefficient}
\end{equation}

At fixed $r>0$, Eqs.~\eqref{eq:u-p-NR} and \eqref{eq:hatr-NR} imply
\begin{equation}
  \begin{aligned}
  \frac{1}{\sqrt{\tilde{r}}}
  &=\frac{1}{\sqrt r}
  \left[1-\epsilon^2P\frac{r-1}{r}+\Order(\epsilon^4)\right],
  \\
  \frac{1}{\mathrm{p}}
  &=1-\epsilon^2P+\Order(\epsilon^4).
  \end{aligned}
  \label{eq:NR-inverse-factors}
\end{equation}
For $\beta=1$, the constants $c_1$ and $\theta$ are held fixed as $\epsilon\to0$. Substituting Eq.~\eqref{eq:NR-inverse-factors} and the shear expansion in Eq.~\eqref{eq:Km-NR-scaling} into Eqs.~\eqref{eq:phi2-a1} and \eqref{eq:S-a1} gives
\begin{equation}
  \begin{aligned}
  \phi_{2,\rel}^{(N_1)}
  &={}\frac{c_1e^{\ii\theta}}{\sqrt r}
  \left[1-\epsilon^2P\frac{r-1}{r}+\Order(\epsilon^4)\right]
  \\
  &\quad\left[\epsilon^2\KNR+\Order(\epsilon^4)\right]
  \\
  &={}\epsilon^2\frac{c_1e^{\ii\theta}}{\sqrt r}\KNR
  +\Order(\epsilon^4)
  \\
  &={}\frac{2\epsilon^2c_1e^{\ii\theta}}{\sqrt r}
  \partial_Z^2f(T,Z)+\Order(\epsilon^4),
  \end{aligned}
  \label{eq:phi2-N1-NR-scaling}
\end{equation}
and
\begin{equation}
  \begin{aligned}
  S_{\rel}^{(N_1)}
  &={}-c_1^2e^{2\ii\theta}
  \left[1-\epsilon^2P+\Order(\epsilon^4)\right]
  \\
  &\quad\left[\epsilon^2\KNR+\Order(\epsilon^4)\right]
  \\
  &={}-\epsilon^2c_1^2e^{2\ii\theta}\KNR
  +\Order(\epsilon^4)
  \\
  &={}-2\epsilon^2c_1^2e^{2\ii\theta}
  \partial_Z^2f(T,Z)+\Order(\epsilon^4).
  \end{aligned}
  \label{eq:S-N1-NR-scaling}
\end{equation}
The remainder estimates refer to the hydrodynamic expansion of the single-copy expressions constructed at first order in the relativistic gradient expansion. Both fields start at order $\epsilon^2$, so their unrescaled limits vanish. Finite, nontrivial leading non-relativistic coefficients are instead defined by
\begin{align}
  \phi_{2,\NR}^{(N_1)}
  &\equiv\lim_{\epsilon\to0}
  \epsilon^{-2}\phi_{2,\rel}^{(N_1)}\notag\\
  &=\frac{2c_1e^{\ii\theta}}{\sqrt r}\partial_Z^2f(T,Z),
  \label{eq:phi2-N1-NR-coefficient}\\
  S_{\NR}^{(N_1)}
  &\equiv\lim_{\epsilon\to0}
  \epsilon^{-2}S_{\rel}^{(N_1)}\notag\\
  &=-2c_1^2e^{2\ii\theta}\partial_Z^2f(T,Z).
  \label{eq:S-N1-NR-coefficient}
\end{align}
On a region of nonzero shear, these coefficients satisfy
\begin{equation}
  \begin{aligned}
  \frac{\bigl(\phi_{2,\NR}^{(N_1)}\bigr)^2}{S_{\NR}^{(N_1)}}
  &=-\frac{\KNR}{r}
  \\
  &=-\frac{2}{r}\partial_Z^2f(T,Z)
  =\Psi_{4,\NR},
  \end{aligned}
  \label{eq:NR-WDC-N1}
\end{equation}
in agreement with the non-relativistic Weyl curvature in Eq.~\eqref{eq:psi4-NR-coefficient}.

For $\beta=0$, the shear dependence is instead carried entirely by the zeroth-copy scalar. Expanding the Maxwell scalar in
Eq.~\eqref{eq:a0-phi-S} gives
\begin{align}
  \phi_{2,\rel}^{(N_0)}
  &=\frac{c_0e^{\ii\theta}}{\sqrt{\tilde{r}}} 
  =\frac{c_0e^{\ii\theta}}{\sqrt r}
  \left[1-\epsilon^2P\frac{r-1}{r}
  +\Order(\epsilon^4)\right],
  \label{eq:phi2-NR-limit}
\end{align}
Similarly, using the pressure expansion in Eq.~\eqref{eq:u-p-NR}
and the shear relations in
Eqs.~\eqref{eq:Km-NR-scaling}-\eqref{eq:KNR},
the zeroth-copy scalar becomes
\begin{align}
  S_{\rel}^{(N_0)}
  &=-\frac{c_0^2e^{2\ii\theta}}
  {\mathrm{p}\Kc^{\rel}}
  =-\epsilon^{-2}
  \frac{c_0^2e^{2\ii\theta}}{\KNR}
  +\Order(\epsilon^0).
  \label{eq:S-NR-scaling}
\end{align}
With \(c_0\) held fixed, the leading contributions to \(\phi_{2,\mathrm{rel}}^{(N_0)}\) and \(\Psi_4^{\mathrm{rel}}\) occur at orders \(\epsilon^0\) and \(\epsilon^2\), respectively. The double copy relation therefore requires an overall factor of \(\epsilon^{-2}\) in \(S_{\mathrm{rel}}^{(N_0)}\). The finite leading coefficient is defined by
\begin{equation}
  S_{\NR}^{(N_0)}
  \equiv\lim_{\epsilon\to0}\epsilon^2S_{\rel}^{(N_0)}.
\end{equation}
Together with Eq.~\eqref{eq:KNR}, this gives
\begin{equation}
  {S_{\NR}^{(N_0)}
  =-\frac{c_0^2e^{2\ii\theta}}
  {2\partial_Z^2f(T,Z)}.}
  \label{eq:SNR}
\end{equation}
The leading non-relativistic Maxwell scalar is defined by
\begin{equation}
  \phi_{2,\NR}^{(N_0)}
  \equiv\lim_{\epsilon\to0}\phi_{2,\rel}^{(N_0)}
  =\frac{c_0e^{\ii\theta}}{\sqrt r}.
\end{equation}
Then
\begin{equation}
  \frac{\bigl(\phi_{2,\NR}^{(N_0)}\bigr)^2}
  {S_{\NR}^{(N_0)}}
  =-\frac{2}{r}\partial_Z^2f(T,Z)
  =\Psi_{4,\NR}.
  \label{eq:NR-WDC}
\end{equation}
Equations~\eqref{eq:NR-WDC-N1} and \eqref{eq:NR-WDC} reproduce the same leading Weyl scalar, but generally involve different Maxwell fields and zeroth-copy scalars. The $\beta=0$ branch recovers the potential-flow single copy of Ref.~\cite{Keeler2020}, whereas the $\beta=1$ branch retains the shear dependence in the leading Maxwell field.

\section{Conclusions}
\label{sec:conclusion}
Using a consistent null tetrad and fixed curvature sign conventions, we have obtained the Weyl scalars, Petrov classification, and Maxwell single copies of the bulk metric dual to a relativistic Rindler fluid at first order in the relativistic gradient expansion. After imposing the first-order fluid equations, the leading curvature is controlled solely by $\mathrm{p}\Omega_{12}$ and $K_{11}-\ii K_{12}$. Detailed Petrov calculations and classifications can be found in Refs.~\cite{Keeler2020,Cai2014,KeelerMonga2025}.

The purely magnetic type D single copy is $F^{(D)}=c_B\calB\dd x \wedge \dd y$. For $\calB=\calB_0$, it is an exact source-free Maxwell solution. For $\calB\sim\Order(\partial)$, it is only a perturbative Maxwell solution valid through first order in gradients. Type N admits the algebraic family in Eqs.~\eqref{eq:phi2-family}-\eqref{eq:S-family}. The choice $\beta=1$ directly encodes the fluid shear in the Maxwell field, whereas $\beta=0$ produces a Maxwell field independent of the potential flow and places all shear information in the zeroth-copy scalar.

Using the non-relativistic hydrodynamic expansion, we determine the leading behavior of both branches. For type D, Eqs.~\eqref{eq:B-NR-scaling} and \eqref{eq:NR-vorticity} give $\calB=\epsilon^2\omega+\Order(\epsilon^4)$. In the purely magnetic phase $\theta=0$, with $c_B$ held fixed, the corresponding curvature, Maxwell scalar, and zeroth-copy scalar behave as
\begin{equation}\begin{aligned}
  &\calB=\epsilon^2\omega+\Order(\epsilon^4),
  \\
  &\Psi_2^{\rel}=\frac{\ii\epsilon^2\omega}{2}
  +\Order(\epsilon^3),
  \\
  &\phi_{1,\rel}^{(D)}=-\frac{\ii c_B\epsilon^2\omega}{2}
  +\Order(\epsilon^4),
  \\
  &S_{\rel}^{(D)}=\frac{\ii c_B^2\epsilon^2\omega}{3}
  +\Order(\epsilon^4).
\end{aligned}
\label{eq:conclusion-D-NR}
\end{equation}
After extracting the coefficient of $\epsilon^2$, we obtain the magnetic field $B_{r,\NR}=c_B\omega$ and the zeroth-copy scalar $S_{\NR}^{(D)}=\ii c_B^2\omega/3$. For constant nonzero $\omega$, these describe a uniform magnetic field normal to the fluid plane and a constant scalar, reproducing the constant-vorticity non-relativistic construction. The Maxwell and zeroth-copy scalars thus have the same leading hydrodynamic order in this branch.

For type N, the two choices of $\beta$ have different leading orders:
\begin{equation}\begin{aligned}
  &\Kc^{\rel}=\epsilon^2\KNR+\Order(\epsilon^4),
  \\
  &\Psi_4^{\rel}=\epsilon^2\Psi_{4,\NR}+\Order(\epsilon^3),
  \\
  &\phi_{2,\rel}^{(N_0)}=\phi_{2,\NR}^{(N_0)}+\Order(\epsilon^2),
  \\
  &S_{\rel}^{(N_0)}=\epsilon^{-2}S_{\NR}^{(N_0)}+\Order(\epsilon^{0}),
  \\
  &\phi_{2,\rel}^{(N_1)}=\epsilon^2\phi_{2,\NR}^{(N_1)}+\Order(\epsilon^4),
  \\
  &S_{\rel}^{(N_1)}=\epsilon^2S_{\NR}^{(N_1)}+\Order(\epsilon^4).
\end{aligned}
\label{eq:conclusion-N-NR}
\end{equation}
We therefore recover the same leading type N Weyl curvature from two distinct allocations of the shear dependence between the Maxwell field and the zeroth-copy scalar. Together with Eq.~\eqref{eq:conclusion-D-NR}, these results establish the non-relativistic limits of the single copies constructed from the relativistic bulk metric at first order in gradients.

Future work may include further analysis of the zeroth-copy scalar wave equation, global branch structure, and possible sources for general $\beta$~\cite{Easson2021}. Other directions include computing the second-order relativistic metric and the corresponding tetrad corrections~\cite{Compere2012,Eling2012,Chirco2011,ElingEntropy2012}. It would also be useful to examine whether the Petrov classification changes~\cite{CaiGB2014} and to compare and analyze the resulting corrections to the Maxwell single copy at second order in the relativistic gradient expansion.

\appendix

\section{Seed metric of Rindler fluid}
\label{app:seed}

Starting from the Rindler metric in ingoing Eddington-Finkelstein coordinates~\cite{Bredberg2012,Keeler2020},
\begin{equation}
  \dd s^2=-\bar r\,\dd\bar\tau^2
  +2\dd\bar\tau\dd\bar r+\dd x_i\dd x^i.
\end{equation}
The induced metric on the surface $\bar r=r_c$ is
\begin{equation}
  \gamma_{ab}\dd \bar x^a\dd \bar x^b
  =-r_c\dd\bar\tau^2+\dd x_i\dd x^i.
\end{equation}
The static unit velocity satisfies
\begin{equation}
  u_a^{(0)}\dd \bar x^a=-\sqrt{r_c}\,\dd\bar\tau,
  \quad
  u_a^{(0)}u_b^{(0)}\dd \bar x^a\dd \bar x^b=r_c\dd\bar\tau^2.
\end{equation}
The original Rindler metric corresponds to the fixed pressure $\mathrm{p}_0=1/\sqrt{r_c}$. An arbitrary constant pressure $\mathrm{p}$ is introduced through the coordinate transformation
\begin{equation}
  \bar r=r+\frac{1}{\mathrm{p}^2}-r_c,
  \quad
  \bar\tau=\sqrt{r_c}\,\mathrm{p}\tau.
  \label{eq:coordinate-transform}
\end{equation}
The time rescaling ensures that the induced metric prescribed by the Dirichlet boundary condition at $r=r_c$ remains unchanged. Substitution gives
\begin{equation}
  \begin{aligned}
  \dd s^2
  ={}&-r_c[1+\mathrm{p}^2(r-r_c)]\dd\tau^2\\
  &+2\sqrt{r_c}\mathrm{p}\dd\tau\dd r+\dd x_i\dd x^i.
  \end{aligned}
\end{equation}
Its covariant form is
\begin{equation}
  \begin{aligned}
  \dd s^2={}&-2\mathrm{p}u_a^{(0)}\dd x^a\dd r\\
  &+[\gamma_{ab}-\mathrm{p}^2(r-r_c)u_a^{(0)}u_b^{(0)}]
  \dd x^a\dd x^b.
  \end{aligned}
\end{equation}
The original horizon $\bar r=0$ is located, in the new coordinates, at
\begin{equation}
  r_H=r_c-\frac{1}{\mathrm{p}^2},
  \quad
  \mathrm{p}=\frac{1}{\sqrt{r_c-r_H}}.
\end{equation}
Thus, $\mathrm{p}$ controls the radial separation between the cutoff surface and the horizon. A boundary Lorentz boost preserving $\gamma_{ab}$, followed by setting $r_c=1$, yields Eq.~\eqref{eq:seed-metric}.

\section{Relativistic gradient expansion}

\label{app:first-order}

After $\mathrm{p}$ and $u_a$ are promoted to slowly varying fields, the relativistic gradient expansion of the bulk metric, with $\delta g_{\mu\nu}$ denoting its first-order correction, takes the form
\begin{equation}
  g_{\mu\nu}=g_{\mu\nu}^{(0)}+\delta g_{\mu\nu}+\Order(\partial^2).
\end{equation}
In radial gauge,
\begin{equation}
  \delta g_{rr}=\delta g_{ra}=0,
\end{equation}
the first-order contribution to the Ricci tensor of the seed metric in Eq.~\eqref{eq:seed-metric}, with $\mathrm{p}$ and $u_a$ promoted to slowly varying fields, is~\cite{Compere2012}
\begin{equation}
  \begin{aligned}
  \widehat R_{ab}
  ={}&\left(D\mathrm{p}+\frac{\mathrm{p}}{2}\Theta\right)u_a u_b\\
  &+\mathrm{p}u_{(a}a_{b)}+u_{(a}\partial_{b)}\mathrm{p},
  \end{aligned}
\end{equation}
with no independent purely transverse tensor projection. The Dirichlet condition requires
\begin{equation}
  \delta g_{ab}\big|_{r=1}=0.
\end{equation}
The general first-order homogeneous structure can therefore be written as
\begin{equation}
  \delta g_{ab}=(1-r)
  [\mathcal{F}u_a u_b+2\mathcal{F}_{(a}u_{b)}].
\end{equation}
Here $\mathcal{F}$ and $\mathcal{F}_a$ are first order in gradients.
On the cutoff surface, the Brown-York stress tensor is
\begin{equation}
  \begin{aligned}
  T_{ab}\big|_{r=1}
  ={}&\Bigl(\mathrm{p}+2D\ln \mathrm{p}
       +\tfrac{\mathcal{F}}{\mathrm{p}}\Bigr)h_{ab}\\
  &+2u_{(a}\Bigl[2a_{b)}
       +\tfrac{\mathcal{F}_{b)}}{\mathrm{p}}\Bigr]-2K_{ab}.
  \end{aligned}
\end{equation}
The pressure is fixed by the isotropic gauge of Ref.~\cite{Compere2012}: derivative corrections contain no independent term proportional to $h_{ab}$. Requiring the coefficient of $h_{ab}$ in $T_{ab}$ to remain $\mathrm{p}$ gives
\begin{equation}
  \mathcal{F}=-2\mathrm{p}D\ln \mathrm{p}=-2D\mathrm{p}.
\end{equation}
The fluid velocity is fixed by the vanishing transverse energy-flux condition, $h_a{}^bT_{bc}u^c=0$~\cite{Compere2012}. This is the Landau-like condition used for Rindler hydrodynamics in Ref.~\cite{Eling2012}. At first order it gives
\begin{equation}
  \mathcal{F}_a=-2\mathrm{p}a_a.
\end{equation}
Substitution yields
\begin{equation}
  \delta g_{ab}
  =2(r-1)[u_a u_bD\mathrm{p}+2\mathrm{p}a_{(a}u_{b)}],
\end{equation}
which is Eq.~\eqref{eq:first-correction} in the main text.

\section{Non-relativistic hydrodynamic expansion}
\label{app:NR-metric}

In the non-relativistic hydrodynamic expansion, the bulk metric dual to a non-relativistic Rindler fluid, used here for comparison with the relativistic gradient expansion, is~\cite{Bredberg2012,Keeler2020,Compere2011}
\begin{equation}\begin{aligned}
\dd s^2={}&-r\dd\tau^2+2\dd\tau\dd r+\dd x^i\dd x^i\\
&-2\left(1-\frac{r}{r_c}\right)v_i\dd x^i\dd\tau
-2\frac{v_i}{r_c}\dd x^i\dd r\\
&+\left(1-\frac{r}{r_c}\right)
\Bigl[(v^2+2P)\dd\tau^2
+\frac{v_i v_j}{r_c}\dd x^i\dd x^j\Bigr]\\
&+\left(\frac{v^2}{r_c}+\frac{2P}{r_c}\right)\dd\tau\dd r
\\
&-\frac{r^2-r_c^2}{r_c}\partial^2v_i\dd x^i\dd\tau
+\Order(\epsilon^3).
\end{aligned}\end{equation}
Under the scaling in Eq.~\eqref{eq:NR-scaling}, the first three terms form the $\Order(\epsilon^0)$ background, the terms linear in $v_i$ form the $\Order(\epsilon)$ velocity perturbation, the terms proportional to $P$ or quadratic in $v_i$ are the $\Order(\epsilon^2)$ pressure and kinetic-energy contributions, and the final derivative term is the viscous contribution. The non-relativistic hydrodynamic expansion and the relativistic gradient expansion assign different orders to these terms. The former can be obtained from the latter through a specific non-relativistic limit~\cite{Compere2012}.

\section*{Acknowledgments}

This work is supported by the National Natural Science Foundation of China (No.12375059).
We acknowledge many helpful discussions with Tucker Manton.

\renewcommand{\bibsection}{

\end{document}
\section*{References}}
\begin{thebibliography}{99}

\bibitem{Bhattacharyya2008}
Sayantani Bhattacharyya, Veronika E Hubeny, Shiraz Minwalla, and Mukund Rangamani,
``Nonlinear Fluid Dynamics from Gravity,''
\emph{JHEP} \textbf{02} (2008) 045,
\href{https://doi.org/10.1088/1126-6708/2008/02/045}{\nolinkurl{doi:10.1088/1126-6708/2008/02/045}},
\href{https://arxiv.org/abs/0712.2456}{arXiv:0712.2456}.

\bibitem{Rangamani2009}
Mukund Rangamani,
``Gravity \& Hydrodynamics: Lectures on the fluid-gravity correspondence,''
\emph{Class. Quantum Grav.} \textbf{26} (2009) 224003,
\href{https://doi.org/10.1088/0264-9381/26/22/224003}{\nolinkurl{doi:10.1088/0264-9381/26/22/224003}},
\href{https://arxiv.org/abs/0905.4352}{arXiv:0905.4352}.

\bibitem{Hubeny2012}
Veronika E. Hubeny, Shiraz Minwalla, and Mukund Rangamani,
``The fluid/gravity correspondence,''
in \emph{Black Holes in Higher Dimensions}, G.~T.~Horowitz (ed.), Cambridge University Press (2012), pp.~348-386,
\href{https://arxiv.org/abs/1107.5780}{arXiv:1107.5780}.

\bibitem{Bredberg2011}
Irene Bredberg, Cynthia Keeler, Vyacheslav Lysov, and Andrew Strominger,
``Wilsonian Approach to Fluid/Gravity Duality,''
\emph{JHEP} \textbf{03} (2011) 141,
\href{https://doi.org/10.1007/JHEP03(2011)141}{\nolinkurl{doi:10.1007/JHEP03(2011)141}},
\href{https://arxiv.org/abs/1006.1902}{arXiv:1006.1902}.

\bibitem{ElingFouxonOz2009}
Christopher Eling, Itzhak Fouxon, and Yaron Oz,
``The Incompressible Navier-Stokes Equations From Black Hole Membrane Dynamics,''
\emph{Phys. Lett. B} \textbf{680} (2009) 496-499,
\href{https://doi.org/10.1016/j.physletb.2009.09.028}{\nolinkurl{doi:10.1016/j.physletb.2009.09.028}},
\href{https://arxiv.org/abs/0905.3638}{arXiv:0905.3638}.

\bibitem{ElingOz2010}
Christopher Eling, and Yaron Oz,
``Relativistic CFT Hydrodynamics from the Membrane Paradigm,''
\emph{JHEP} \textbf{02} (2010) 069,
\href{https://doi.org/10.1007/JHEP02(2010)069}{\nolinkurl{doi:10.1007/JHEP02(2010)069}},
\href{https://arxiv.org/abs/0906.4999}{arXiv:0906.4999}.

\bibitem{LysovStrominger2011}
Vyacheslav Lysov, and Andrew Strominger,
``From Petrov-Einstein to Navier-Stokes,''
(2011),
\href{https://arxiv.org/abs/1104.5502}{arXiv:1104.5502}.

\bibitem{Huang2011}
Tai-Zhuo Huang, Yi Ling, Wen-Jian Pan, Yu Tian, and Xiao-Ning Wu,
``From Petrov-Einstein to Navier-Stokes in Spatially Curved Spacetime,''
\emph{JHEP} \textbf{10} (2011) 079,
\href{https://doi.org/10.1007/JHEP10(2011)079}{\nolinkurl{doi:10.1007/JHEP10(2011)079}},
\href{https://arxiv.org/abs/1107.1464}{arXiv:1107.1464}.

\bibitem{Bredberg2012}
I.~Bredberg, C.~Keeler, V.~Lysov, and A.~Strominger,
``From Navier-Stokes to Einstein,''
\emph{JHEP} \textbf{07} (2012) 146,
\href{https://doi.org/10.1007/JHEP07(2012)146}{\nolinkurl{doi:10.1007/JHEP07(2012)146}},
\href{https://arxiv.org/abs/1101.2451}{arXiv:1101.2451}.

\bibitem{Compere2012}
G.~Comp\`ere, P.~McFadden, K.~Skenderis, and M.~Taylor,
``The relativistic fluid dual to vacuum Einstein gravity,''
\emph{JHEP} \textbf{03} (2012) 076,
\href{https://doi.org/10.1007/JHEP03(2012)076}{\nolinkurl{doi:10.1007/JHEP03(2012)076}},
\href{https://arxiv.org/abs/1201.2678}{arXiv:1201.2678}.

\bibitem{Eling2012}
C.~Eling, A.~Meyer, and Y.~Oz,
``The Relativistic Rindler Hydrodynamics,''
\emph{JHEP} \textbf{05} (2012) 116,
\href{https://doi.org/10.1007/JHEP05(2012)116}{\nolinkurl{doi:10.1007/JHEP05(2012)116}},
\href{https://arxiv.org/abs/1201.2705}{arXiv:1201.2705}.

\bibitem{Compere2011}
Geoffrey Comp\`ere, Paul McFadden, Kostas Skenderis, and Marika Taylor,
``The holographic fluid dual to vacuum Einstein gravity,''
\emph{JHEP} \textbf{07} (2011) 050,
\href{https://doi.org/10.1007/JHEP07(2011)050}{\nolinkurl{doi:10.1007/JHEP07(2011)050}},
\href{https://arxiv.org/abs/1103.3022}{arXiv:1103.3022}.

\bibitem{Keeler2020}
C.~Keeler, T.~Manton, and N.~Monga,
``From Navier-Stokes to Maxwell via Einstein,''
\emph{JHEP} \textbf{08} (2020) 147,
\href{https://doi.org/10.1007/JHEP08(2020)147}{\nolinkurl{doi:10.1007/JHEP08(2020)147}},
\href{https://arxiv.org/abs/2005.04242}{arXiv:2005.04242}.

\bibitem{Bern2008}
Z. Bern, J. J. M. Carrasco, and H. Johansson,
``New Relations for Gauge-Theory Amplitudes,''
\emph{Phys. Rev. D} \textbf{78} (2008) 085011,
\href{https://doi.org/10.1103/PhysRevD.78.085011}{\nolinkurl{doi:10.1103/PhysRevD.78.085011}},
\href{https://arxiv.org/abs/0805.3993}{arXiv:0805.3993}.

\bibitem{Bern2010}
Z.~Bern, J.~J.~M.~Carrasco, and H.~Johansson,
``Perturbative Quantum Gravity as a Double Copy of Gauge Theory,''
\emph{Phys. Rev. Lett.} \textbf{105} (2010) 061602,
\href{https://doi.org/10.1103/PhysRevLett.105.061602}{\nolinkurl{doi:10.1103/PhysRevLett.105.061602}},
\href{https://arxiv.org/abs/1004.0476}{arXiv:1004.0476}.

\bibitem{Monteiro2014}
Ricardo Monteiro, Donal O'Connell, and Chris D. White,
``Black holes and the double copy,''
\emph{JHEP} \textbf{12} (2014) 056,
\href{https://doi.org/10.1007/JHEP12(2014)056}{\nolinkurl{doi:10.1007/JHEP12(2014)056}},
\href{https://arxiv.org/abs/1410.0239}{arXiv:1410.0239}.

\bibitem{Luna2019}
A.~Luna, R.~Monteiro, I.~Nicholson, and D.~O'Connell,
``Type D Spacetimes and the Weyl Double Copy,''
\emph{Class. Quantum Grav.} \textbf{36} (2019) 065003,
\href{https://doi.org/10.1088/1361-6382/ab03e6}{\nolinkurl{doi:10.1088/1361-6382/ab03e6}},
\href{https://arxiv.org/abs/1810.08183}{arXiv:1810.08183}.

\bibitem{White2021}
Chris D. White,
``Twistorial Foundation for the Classical Double Copy,''
\emph{Phys. Rev. Lett.} \textbf{126} (2021) 061602,
\href{https://doi.org/10.1103/PhysRevLett.126.061602}{\nolinkurl{doi:10.1103/PhysRevLett.126.061602}},
\href{https://arxiv.org/abs/2012.02479}{arXiv:2012.02479}.

\bibitem{Godazgar2021}
H.~Godazgar, M.~Godazgar, R.~Monteiro, D.~Peinador Veiga, and C.~N.~Pope,
``Weyl Double Copy for Gravitational Waves,''
\emph{Phys. Rev. Lett.} \textbf{126} (2021) 101103,
\href{https://doi.org/10.1103/PhysRevLett.126.101103}{\nolinkurl{doi:10.1103/PhysRevLett.126.101103}},
\href{https://arxiv.org/abs/2010.02925}{arXiv:2010.02925}.

\bibitem{Easson2021}
Damien A. Easson, Tucker Manton, and Andrew Svesko,
``Sources in the Weyl double copy,''
\emph{Phys. Rev. Lett.} \textbf{127} (2021) 271101,
\href{https://doi.org/10.1103/PhysRevLett.127.271101}{\nolinkurl{doi:10.1103/PhysRevLett.127.271101}},
\href{https://arxiv.org/abs/2110.02293}{arXiv:2110.02293}.

\bibitem{Easson2023}
D.~A.~Easson, T.~Manton, and A.~Svesko,
``Einstein-Maxwell theory and the Weyl double copy,''
\emph{Phys. Rev. D} \textbf{107} (2023) 044063,
\href{https://doi.org/10.1103/PhysRevD.107.044063}{\nolinkurl{doi:10.1103/PhysRevD.107.044063}},
\href{https://arxiv.org/abs/2210.16339}{arXiv:2210.16339}.

\bibitem{Adamo2022Snowmass}
T.~Adamo, J.~J.~M.~Carrasco, M.~Carrillo-Gonz\'alez, M.~Chiodaroli, H.~Elvang, H.~Johansson, D.~O'Connell, R.~Roiban, and O.~Schlotterer,
``Snowmass White Paper: the Double Copy and its Applications,''
(2022),
\href{https://arxiv.org/abs/2204.06547}{arXiv:2204.06547}.

\bibitem{CheungMangan2020}
Clifford Cheung, and James Mangan,
``Scattering Amplitudes and the Navier-Stokes Equation,''
(2020),
\href{https://arxiv.org/abs/2010.15970}{arXiv:2010.15970}.

\bibitem{KeelerMonga2025}
C.~Keeler and N.~Monga,
``Type-II Spacetimes and the Double Copy for Fluids Metrics,''
\emph{Phys. Rev. D} \textbf{112} (2025) 044045,
\href{https://doi.org/10.1103/6h4v-3l3r}{\nolinkurl{doi:10.1103/6h4v-3l3r}},
\href{https://arxiv.org/abs/2404.03195}{arXiv:2404.03195}.

\bibitem{MeyerOz2013}
Adiel Meyer, and Yaron Oz,
``Constraints on Rindler Hydrodynamics,''
\emph{JHEP} \textbf{07} (2013) 090,
\href{https://doi.org/10.1007/JHEP07(2013)090}{\nolinkurl{doi:10.1007/JHEP07(2013)090}},
\href{https://arxiv.org/abs/1304.6305}{arXiv:1304.6305}.

\bibitem{BrownYork1993}
J. David Brown, and James W. York, Jr.,
``Quasilocal Energy and Conserved Charges Derived from the Gravitational Action,''
\emph{Phys. Rev. D} \textbf{47} (1993) 1407-1419,
\href{https://doi.org/10.1103/PhysRevD.47.1407}{\nolinkurl{doi:10.1103/PhysRevD.47.1407}},
\href{https://arxiv.org/abs/gr-qc/9209012}{arXiv:gr-qc/9209012}.

\bibitem{Elor2020}
G.~Elor, K.~Farnsworth, M.~L.~Graesser, and G.~Herczeg,
``The Newman-Penrose Map and the Classical Double Copy,''
\emph{JHEP} \textbf{12} (2020) 121,
\href{https://doi.org/10.1007/JHEP12(2020)121}{\nolinkurl{doi:10.1007/JHEP12(2020)121}},
\href{https://arxiv.org/abs/2006.08630}{arXiv:2006.08630}.

\bibitem{NewmanPenrose1962}
E.~Newman and R.~Penrose,
``An Approach to Gravitational Radiation by a Method of Spin Coefficients,''
\emph{J. Math. Phys.} \textbf{3} (1962) 566-578,
\href{https://doi.org/10.1063/1.1724257}{\nolinkurl{doi:10.1063/1.1724257}}.

\bibitem{Cai2014}
R.-G.~Cai, Q.~Yang, and Y.-L.~Zhang,
``Petrov type I Spacetime and Dual Relativistic Fluids,''
\emph{Phys. Rev. D} \textbf{90} (2014) 041901(R),
\href{https://doi.org/10.1103/PhysRevD.90.041901}{\nolinkurl{doi:10.1103/PhysRevD.90.041901}},
\href{https://arxiv.org/abs/1401.7792}{arXiv:1401.7792}.

\bibitem{CaiLiYangZhang2013}
Rong-Gen Cai, Li Li, Qing Yang, and Yun-Long Zhang,
``Petrov type I Condition and Dual Fluid Dynamics,''
\emph{JHEP} \textbf{04} (2013) 118,
\href{https://doi.org/10.1007/JHEP04(2013)118}{\nolinkurl{doi:10.1007/JHEP04(2013)118}},
\href{https://arxiv.org/abs/1302.2016}{arXiv:1302.2016}.

\bibitem{Chacon2021}
E.~Chac\'on, S.~Nagy, and C.~D.~White,
``The Weyl double copy from twistor space,''
\emph{JHEP} \textbf{05} (2021) 239,
\href{https://doi.org/10.1007/JHEP05(2021)239}{\nolinkurl{doi:10.1007/JHEP05(2021)239}},
\href{https://arxiv.org/abs/2103.16441}{arXiv:2103.16441}.

\bibitem{Luna2015TaubNUT}
A.~Luna, R.~Monteiro, D.~O'Connell, and C.~D.~White,
``The classical double copy for Taub-NUT spacetime,''
\emph{Phys. Lett. B} \textbf{750} (2015) 272-277,
\href{https://doi.org/10.1016/j.physletb.2015.09.021}{\nolinkurl{doi:10.1016/j.physletb.2015.09.021}},
\href{https://arxiv.org/abs/1507.01869}{arXiv:1507.01869}.

\bibitem{BahjatAbbas2017}
N.~Bahjat-Abbas, A.~Luna, and C.~D.~White,
``The Kerr-Schild double copy in curved spacetime,''
\emph{JHEP} \textbf{12} (2017) 004,
\href{https://doi.org/10.1007/JHEP12(2017)004}{\nolinkurl{doi:10.1007/JHEP12(2017)004}},
\href{https://arxiv.org/abs/1710.01953}{arXiv:1710.01953}.

\bibitem{Bhattacharyya2009}
Sayantani Bhattacharyya, Shiraz Minwalla, and Spenta R. Wadia,
``The Incompressible Non-Relativistic Navier-Stokes Equation from Gravity,''
\emph{JHEP} \textbf{08} (2009) 059,
\href{https://doi.org/10.1088/1126-6708/2009/08/059}{\nolinkurl{doi:10.1088/1126-6708/2009/08/059}},
\href{https://arxiv.org/abs/0810.1545}{arXiv:0810.1545}.

\bibitem{Chirco2011}
Goffredo Chirco, Christopher Eling, and Stefano Liberati,
``Higher Curvature Gravity and the Holographic fluid dual to flat spacetime,''
\emph{JHEP} \textbf{08} (2011) 009,
\href{https://doi.org/10.1007/JHEP08(2011)009}{\nolinkurl{doi:10.1007/JHEP08(2011)009}},
\href{https://arxiv.org/abs/1105.4482}{arXiv:1105.4482}.

\bibitem{ElingEntropy2012}
Christopher Eling, Adiel Meyer, and Yaron Oz,
``Local Entropy Current in Higher Curvature Gravity and Rindler Hydrodynamics,''
\emph{JHEP} \textbf{08} (2012) 088,
\href{https://doi.org/10.1007/JHEP08(2012)088}{\nolinkurl{doi:10.1007/JHEP08(2012)088}},
\href{https://arxiv.org/abs/1205.4249}{arXiv:1205.4249}.

\bibitem{CaiGB2014}
Rong-Gen Cai, Qing Yang, and Yun-Long Zhang,
``Petrov type I Condition and Rindler Fluid in Vacuum Einstein-Gauss-Bonnet Gravity,''
\emph{JHEP} \textbf{12} (2014) 147,
\href{https://doi.org/10.1007/JHEP12(2014)147}{\nolinkurl{doi:10.1007/JHEP12(2014)147}},
\href{https://arxiv.org/abs/1408.6488}{arXiv:1408.6488}.

\bibitem{Godazgar2021Asymptotic}
H.~Godazgar, M.~Godazgar, R.~Monteiro, D.~Peinador Veiga, and C.~N.~Pope,
``Asymptotic Weyl Double Copy,''
\emph{JHEP} \textbf{11} (2021) 126,
\href{https://doi.org/10.1007/JHEP11(2021)126}{\nolinkurl{doi:10.1007/JHEP11(2021)126}},
\href{https://arxiv.org/abs/2109.07866}{arXiv:2109.07866}.

\bibitem{AdamoKol2022}
T.~Adamo and U.~Kol,
``Classical double copy at null infinity,''
\emph{Class. Quantum Grav.} \textbf{39} (2022) 105007,
\href{https://doi.org/10.1088/1361-6382/ac635e}{\nolinkurl{doi:10.1088/1361-6382/ac635e}},
\href{https://arxiv.org/abs/2109.07832}{arXiv:2109.07832}.

\end{thebibliography}
